# TraianProt: a user-friendly R shiny application for wide format proteomics data downstream analysis

*Samuel de la Cámara Fuentes[1], Dolores Gutiérrez-Blázquez[1], María Luisa Hernáez[1] and Concha Gil[1,2].*

1. Proteomics Unit, Faculty of Pharmacy, Complutense University of Madrid, Madrid, Spain.
2. Microbiology and Parasitology Department, Faculty of Pharmacy, Complutense University of Madrid, Madrid, Spain.

**Abstract**

**Background.** Mass spectrometry–based proteomics enables large-scale identification, quantification, and characterization of proteins, post-translational modifications, and dynamic expression changes, providing critical insights into biological processes and disease mechanisms. However, downstream data analysis remains a significant challenge due to the heterogeneity of datasets, diverse computational platforms, and varying output formats, often requiring advanced expertise. To overcome these barriers, user-friendly, web-based solutions have emerged as essential tools, facilitating accessible, customizable, and reproducible analysis while accelerating discovery in proteomics research.

**Results.** We present TraianProt, a web-based, user-friendly proteomics data analysis platform that enables the analysis of both label-free and labeled data from Data-Dependent or Data-Independent Acquisition mass spectrometry mode, supporting MaxQuant, MSFragger, DIA-NN, ProteoScape, and Proteome Discoverer output formats. TraianProt provides a comprehensive suite of stepwise downstream analysis modules, which includes data filtering, normalization procedures, and missing value imputation strategies. The platform also incorporates robust statistical frameworks for differential expression testing, with peptide-spectrum match level correction, thereby enhancing the reliability of biological interpretations. Beyond quantitative comparisons, TraianProt integrates enrichment analyses to contextualize proteomic changes within functional pathways and interaction network analyses allowing users to extract biological insights from proteomic data without any programming skills.

**Conclusions.** TraianProt serves as a comprehensive data analysis tool for proteomic studies, capable of analyzing outputs from main search algorithms without manual reformatting. The tool is open-source and freely available to the community. An interactive web server is accessible at https://samueldelacamara.shinyapps.io/TraianProt/. Furthermore, the complete source code, user tutorials, and an installable R package are all available at the project's GitHub repository: https://github.com/SamueldelaCamaraFuentes/TraianProt.

Corresponding author: sdelacam@ucm.es

## 1. Background

Mass spectrometry-based proteomics is a high-powered tool that enables the qualitative and quantitative characterization of thousands of peptides and proteins in biological samples across various experimental conditions. This approach plays a central role in biology, as it not only facilitates the identification of proteins but also allows researchers to quantify their abundance, monitor post-translational modifications, and assess dynamic changes in protein expression. As such, proteomics has become a key strategy for gaining insight into complex biological processes, identifying novel biomarkers disease, and studying intricate molecular mechanisms such as cell signaling pathways and protein–protein interactions. Despite its transformative potential, the downstream analysis of proteomics data remains a major bottleneck. The diversity of proteomics data processing tools, each using distinct algorithms and output formats, complicates downstream analysis and requires advanced computational expertise for accurate biological interpretation.

In addition to computational challenges, the inherent complexity and heterogeneity of proteomic datasets further complicate analysis. Addressing specific biological questions often requires customizable and flexible analytical approaches, which highlights the need for accessible solutions. Therefore, the development of intuitive, user-friendly web platforms has become an important advancement for empowering scientists in the efficient exploration, visualization, and interpretation of proteomics data. By lowering the computational barrier, such tools not only facilitate data-driven discovery but also contribute to advancing life science research. Bioinformaticians and computational biologists have developed several web platforms for the analysis of large-scale proteomics datasets.

The first example of these efforts is Perseus [1], a popular tool for data analysis which provides a graphical user interface for the analysis of both label-based and label-free methods. However, its applicability is restricted, as it only supports the MaxQuant [2] output format, which limits the range of analysis possibilities for proteomic data, as manual formatting is required for analyzing data from other search engine outputs. Another software is ProVision [3], which provides a broad set of functionalities through an intuitive and user-friendly interface. It facilitates fast exploratory and statistical data analysis by providing sensible default parameters, thereby lowering the entry barrier for users without advanced computational expertise. Nevertheless, ProVision also relies exclusively on MaxQuant output files for downstream analysis, which restricts its usability. Finally, it is necessary to highlight Fragpipe Analyst [4] which enables the automatic analysis of label free data of MSFragger [5] data output-format which is regarded as a widely adopted resource in the proteomics field, By contrast, similar to Perseus and ProVision, this software supports only the output generated by the various FragPipe workflows Consequently, users who wish to analyze data from other proteomics platforms or to adapt protein group datasets may encounter significant difficulties, particularly if they lack the necessary computational expertise which may lead to a considerable burden in terms of time, effort and the risk of generating suboptimal results. Furthermore, these established software tools exhibit more limitations, for example, they might lack integrated, broad-spectrum functional annotation, which restricts analysis to common model organisms. Another issue is that they frequently lack essential data-filtering capabilities, such as options for unique peptide-centric filtering which is only implemented in Provision or PSM-level correction. These capabilities are necessary for the identification of high-confidence biomarker candidates.

This analytical gap persists even in other established tools. Amica [6] and ProStaR [7], for instance, lack capabilities for stringent unique peptide filtering and PSM-based variance correction. Conversely, while MCQR [8] commendably integrates retention time data, it lacks critical downstream modules for both functional annotation and interaction network analysis.

Here we present TraianProt, a web platform that provides comprehensive tools for the analysis of quantitative proteomics data which enables the analysis of label-free and labeled data as it is the case in tandem mass tag (TMT) and Stable Isotope Labeling with Amino Acid in Cell Culture (SILAC) from Data-Dependent Acquisition mass spectrometry (DDA-MS) or Data-Independent Acquisition mass spectrometry (DIA-MS) mode, supporting MaxQuant [2], MSFragger [5] , DIA-NN [9] , ProteoScape (Bruker, Massachusetts, USA) and Proteome Discoverer (Thermo Scientific, Waltham, MA, USA) output formats. TraianProt performs a complete downstream analysis of proteomic data covering several postprocessing steps, such as filtering, normalization and imputation methods allocated in the “Pre-processing module”, hypothesis testing in the “Differential analysis module”, functional enrichment analysis from the “Functional Analysis module” and protein-protein interactions using STRING in the “Interaction analysis module”. Results of all these modules generate publication-ready plots. Finally, TraianProt is built in the R-Shiny web framework, which confers its reactive nature allowing the user to make changes that are propagated through the platform. This helps users to find the most accurate configuration of parameters for the analysis.

In light of this, two examples of proteomic downstream analysis will be addressed to demonstrate the functionalities of the TraianProt application. The first example focuses on a DDA experiment in which the raw files were processed with MaxQuant. The second example involves a DIA case study where the raw files were analyzed with the DIA-NN software. Together, these case studies provide complementary scenarios that demonstrate how TraianProt can accommodate different acquisition strategies and software outputs.

## 2. Implementation

TraianProt is implemented entirely in the R programming language (v4.2) and is built upon the Shiny framework (v1.11.1) for its graphical user interface.

The tool is open-source and freely available to the community in several formats. The interactive web server is hosted at https://samueldelacamara.shinyapps.io/TraianProt/. In addition, a standalone R package, the complete source code, and user tutorials are all available at https://github.com/SamueldelaCamaraFuentes/TraianProt.

### 2.1 Field of application

TraianProt is designed to support both label-free and label-based proteomics workflows. Label-free quantification uses spectral counts and ion intensities to estimate protein abundance without isotopic labelling. In contrast, label-based methods employ stable isotope tagging to enable multiplexed quantification. Examples of the latter include TMT, which uses reporter ion intensities, and SILAC, which distinguishes samples by incorporating isotopic variants during protein synthesis. The platform can also handle data acquired through data-dependent (DDA) and data-independent acquisition (DIA) strategies. In DDA, precursor ions are selected for fragmentation based on intensity, while in DIA, all ions within defined m/z ranges are fragmented simultaneously to achieve comprehensive peptide coverage [10, 11].

### 2.2 Architecture and functional modules

The architecture of TraianProt covers 6 different modules: preprocessing, quality control, differential analysis, differential analysis plots, functional analysis and interaction analysis module (Figure 1).

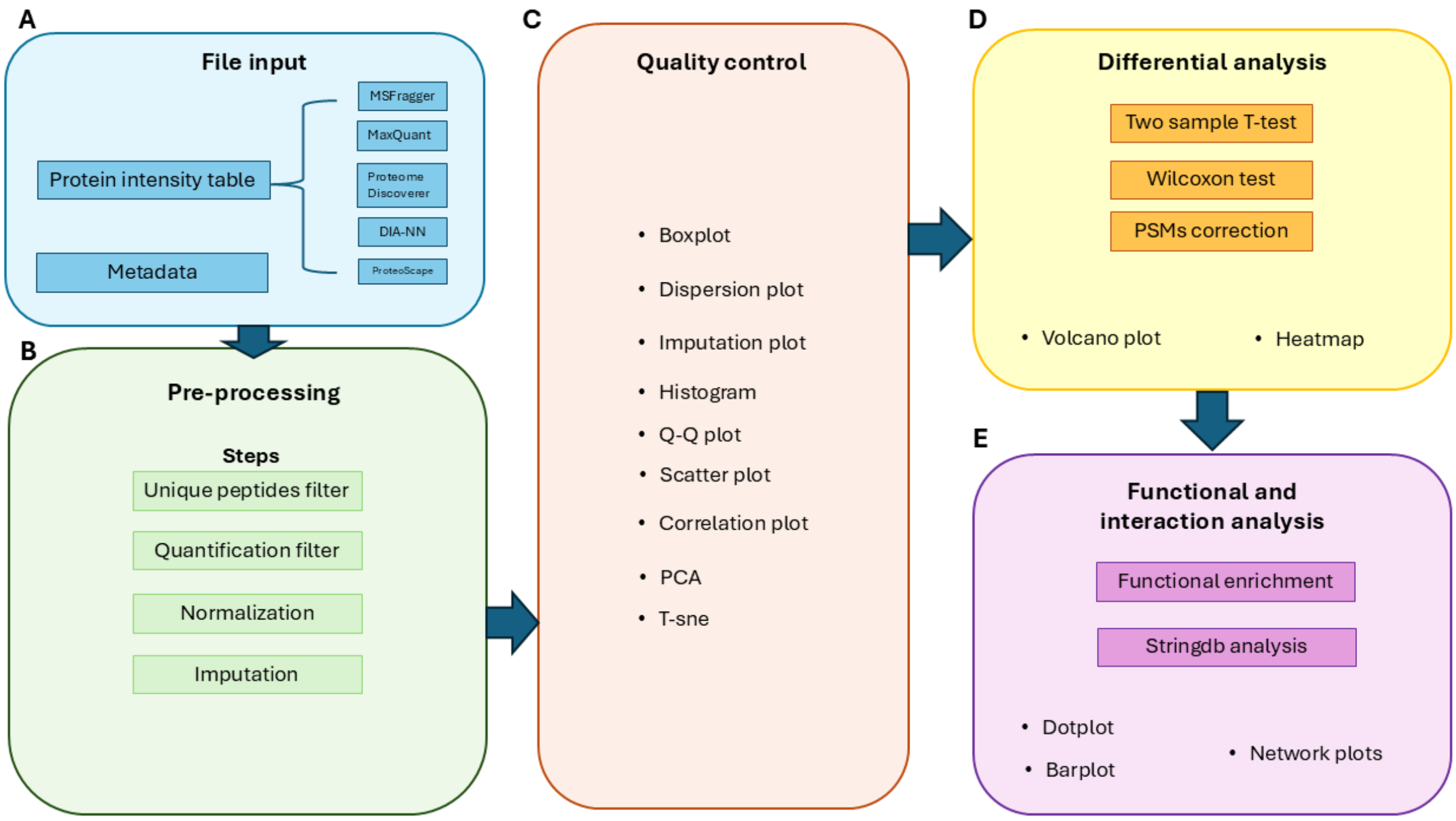

**Figure 1** Overview of the architecture and functional modules of TraianProt. A) Data import supporting protein group tables format from MSFragger, MaxQuant, Proteome Discoverer, DIA-NN and ProteoScape. B) Pre-processing steps involving filtering, normalization and imputation. C) Quality control. D) Differential analysis. E) Functional and interaction analysis.

### 2.2.1 Data import

Datasets are uploaded through the "Pre-processing" module in the "File input" section. Both a dataset containing protein groups along with their intensity values and a metadata file containing the experimental design are required. A detailed description of the required files can be found in the "File input section" in Supplementary file 1, this includes all the necessary instructions for using the software. The proteomic protein data and metadata are expected to be in tab-delimited text file format (except Proteome Discoverer protein data which is expected to be in its standard .xlsx output format) with columns such as Protein Identifiers, Peptide-Spectrum Matches (PSMs), matching the experimental mass spectra to theoretical spectra), Peptides (number of peptides that can be mapped to a protein) and Intensity values (sum of intensities of the peptides mapped to a protein) in the proteomic protein data table and columns such as sample name, group and intensity value names in the metadata file.

### 2.2.2 Data pre-processing

No matter which platform generated the dataset, once data has been submitted, it will undergo a pre-processing step involving the transformation of sample intensities into a logarithmic base 2 scale. A set of options can optionally be applied to data such as the unique peptides filter. A unique peptide is a peptide that is only found in one protein group and is not shared with any other protein group. To ensure only high-confidence protein quantifications are used for downstream statistical analysis, the unique peptides filter is applied where protein groups are filtered based on both the number of unique peptides and the consistency of detection. Specifically, a protein group is retained for analysis only if it was quantified by a user-defined minimum of unique peptides (e.g.,≥ 2) in a user-defined minimum proportion of samples e.g.,≥ 50%) in at least one experimental condition. This is an important step for selecting robust proteins thereby increasing the statistical power and reliability of the subsequent differential expression analysis and biomarker nomination. In the case of DIA-NN output results, the web-server implementation does not support the unique peptide filter, since the required information is extracted from the “report.tsv” file, which typically exceeds the maximum upload size permitted in a Shiny application. This feature, however, is fully supported when TraianProt is executed locally through RStudio.

Additionally, a quantification-based filter is also applied to retain protein groups identified in a minimal proportion of samples. The intensity values of the filtered protein groups can be selected for normalization. Several normalization methods are available including mean, median and variance stabilization normalization (VSN). The selected method will be applied by entering it into the 'Normalization method' prompt. Finally, two different methods of imputing missing values can be chosen: i) “Normal distribution” which commits the imputation considering a normal distribution with a user specified downshift and width, ii) “K Nearest Neighbors” which uses the KNN algorithm for imputation. Please note that it is possible not to select neither an imputation method and handle missing values in steps to be taken downstream nor a normalization method.

The quality control module comprises a series of plots that describe the nature of data. These include boxplot, Q-Q plots, histograms, pre- and post-imputation plots, dispersion plots, scatter, correlation and principal component analysis plots. All these plots can be downloaded in publication-ready TIFF format. This module shows whether the changes in the pre-processing step have been applied correctly and suggests the best parameters to select. The reactive nature of the entire application enables changes to propagate automatically through subsequent steps in the app.

**2.2.3 Differential analysis and visualization**

The Differential Analysis module facilitates both the examination and visualization of relative quantitative proteomics datasets. At the outset, users can choose to perform either a paired or an unpaired two-sample t-test. To this end, users can select between the base R Student´s t-Test function, the T-Test function approach from the limma [12] package or the Wilcoxon test function from base R depending on distribution followed by data. If the limma approach is selected, a PSMs correction using the DEqMS [13] package can be applied to achieve a more accurate variance estimation. This method accounts for the inherent dependence of protein variance on the number of PSMs or peptides used for quantification when assessing differential protein expression, which allows for a more accurate data-dependent estimation of protein variance and inclusion of single peptide identifications without increasing false discoveries.

In the next step users can specify common thresholds to subset common protein groups of interest, such as $\log_2$ fold change threshold, thresholds on *p*-value or adjusted *p*-value threshold. In differential expression proteomic analyses, there are proteins that we call 'exclusive proteins'. These are proteins that exhibit absence or presence and are identified exclusively in one of the studied conditions. They are quite interesting because their expression is only quantified in one of the two conditions under study. These proteins are labelled as 'unique' due to their absence in one condition and their presence in more than 50% of samples in the other condition (i.e. they must display intensity values in more than 50% of samples). In addition, to be included in the list of differentially abundant proteins, the values of $\log_2$FC and p-value are added to the exclusive proteins. For $\log_2$FC, the value is computed from the lowest value in the distribution of ratios for down-regulated proteins, minus two, and the highest value in the distribution of ratios for up-regulated proteins, plus two. For the *p*-value /adjusted

*p*-value is computed from the lowest value in the distribution divided by a factor of 100. In this case, the integration of exclusive proteins into the list of differentially abundant proteins is assessed, and these proteins will be considered in the next steps of the analysis. However, if the user does not want to include these 'common proteins', they can select 'Choose a way to proceed' in the prompt.

The statistical analysis is fully reactive as well, which allows statistical changes to propagate to the next sections. A resulting data frame composed of differentially abundant proteins including non-differentially abundant proteins and containing several statistical metrics such as fold change values, *p*-value/ adjusted *p*-value is presented in a tabular format which can be downloaded as a csv file for subsequent analysis.

Differentially abundant proteins are visualized as volcano plot and heatmap plot in the Differential analysis plot module. The volcano plot enables the visualization of data within an experimental setup, users can customize the points size and the title. Moreover, two heatmaps are displayed, the first one displays the identified proteins and the second one those proteins that exhibited differential relative abundances. Finally, there is a protein intensity plot which displays log2 intensity of a protein across the samples of every condition, once its protein id is entered in the application. These plots can be exported in various file formats, including publication-ready TIFF format.

#### 2.2.4 Functional analysis

The functional analysis module covers the retrieval of statistically significant enriched terms from public domains and their integration with the proteins identified in the proteomics data set. It is performed using g:Profiler [14] and assisted in the Functional analysis module where users can select the organism of the experiment, introducing the organism id as it is explained in the "Functional analysis section" of Supplementary file 1. The output of g:Profiler constituted by a data table containing biological terms and pathways is displayed in various plots including a dot plot, a bar plot of terms and a Manhattan plot. Parameters such as the number of terms to display in the plot, the font size and the title can be customized. These plots can be also exported in various file formats, including publication-ready TIFF format. Additionally, an output table containing data sources, term names, p-values, and other relevant attributes is displayed and can be downloaded as a xlsx file.

#### 2.2.5 Interaction analysis

The Interaction analysis module performs a protein-protein interaction analysis using the STRINGdb [15] R package where users need to specify the NCBI taxonomy identifier of the organism on which the experiment has been performed. STRING aggregates data from many different sources in a way that results in edge confidence scores between 0 and 1. By default STRING will show all interaction with a medium confidence (above 0.4), any interactions with scores below this threshold will not be shown.

STRING currently has two different network types: physical interactions and functional associations. In the case of functional associations, the edges indicate if the proteins are functionally related, such as being part of the same pathway, while for physical interactions, it is the likelihood of the proteins to be in the same protein complex. A network illustrating both direct (physical) and indirect (functional) associations between differentially abundant proteins is displayed. One network is shown for proteins with increased abundance, and another for proteins with decreased abundance. Finally, a table summarizing important metrics for graph analysis such as (degree, betweenness, closeness…) is attached.

## 3. Results

### 3.1 User interface

The TraianProt web application is created using the R Shiny package (v1.11.1) and designed with packages such as shinywidgets (v0.9), shinydashboard (v0.7.3) and shinyjs (v2.1.0) with the aim of creating a user-friendly experience. The shiny framework allows users to make changes that will be propagated throughout subsequent steps in the software. The interface can be divided into seven parts, each being referred to a module: home, Preprocessing, Quality control, Differential analysis, Differential analysis plots, Functional analysis and Interaction analysis module.

### 3.2 Use study 1: MaxQuant dataset

The example data was taken from a label-free Data-Dependent Acquisition proteomic analysis focusing on Prn1[16], a protein with an unknown function in *C. albicans* similar to mammalian Pirin. In this study the effect of $H_2O_2$ in the presence or absence of Prn1 is studied using a null *C. albicans prn1*Δ mutant and the corresponding wild-type strain SN250. It was concluded that it is a relevant component of the oxidative stress response in *C. albicans*. This includes raw files for four groups containing 4 samples per group deposited in the ProteomeXchange Consortium via Proteomics Identifications Database (PRIDE) partner repository [17] with the dataset identifier PXD040804.

Raw files were re-processed using MaxQuant (version 2.1.4.0) with similar parameters as the results presented in [16] which were originally analyzed using the Proteome Discoverer (version 2.4) pipeline. Here we present an alternative approach to demonstrate the potential of TraianProt in analysing wide-format data. The Candida Genome Database (release 2020_06, 6209 sequences) (CGD) was used. Search parameters included Carbamidomethylation of cysteines as fixed modification, oxidation of methionine and N-terminal acetylation as variable modifications and Trypsin/P as proteolytic enzyme with a maximum of 2 missed cleavages allowed. The precursor mass tolerance was 10 ppm and the fragment mass tolerance was 0.02 Da.

The proteomic analysis enabled the quantification of 1748 *C. albicans* proteins in the SN250 $H_2O_2$ vs SN250 control comparison. The intensity values of the quantified proteins were transformed into a logarithmic scale. In addition, both decoy and contaminant proteins were removed. Only common protein groups (with intensity values in at least 50 % of the samples from both conditions) are included in the analysis. Furthermore, a protein group is retained when at least one unique peptide has been identified in a proportion of 50 % of the samples in at least one of the two groups being compared. The normalization method applied was median normalization; protein intensities were normalized by subtracting the median for each biological replicate. No imputation method was selected, missing values will be handled in steps taken downstream (Figure 2).

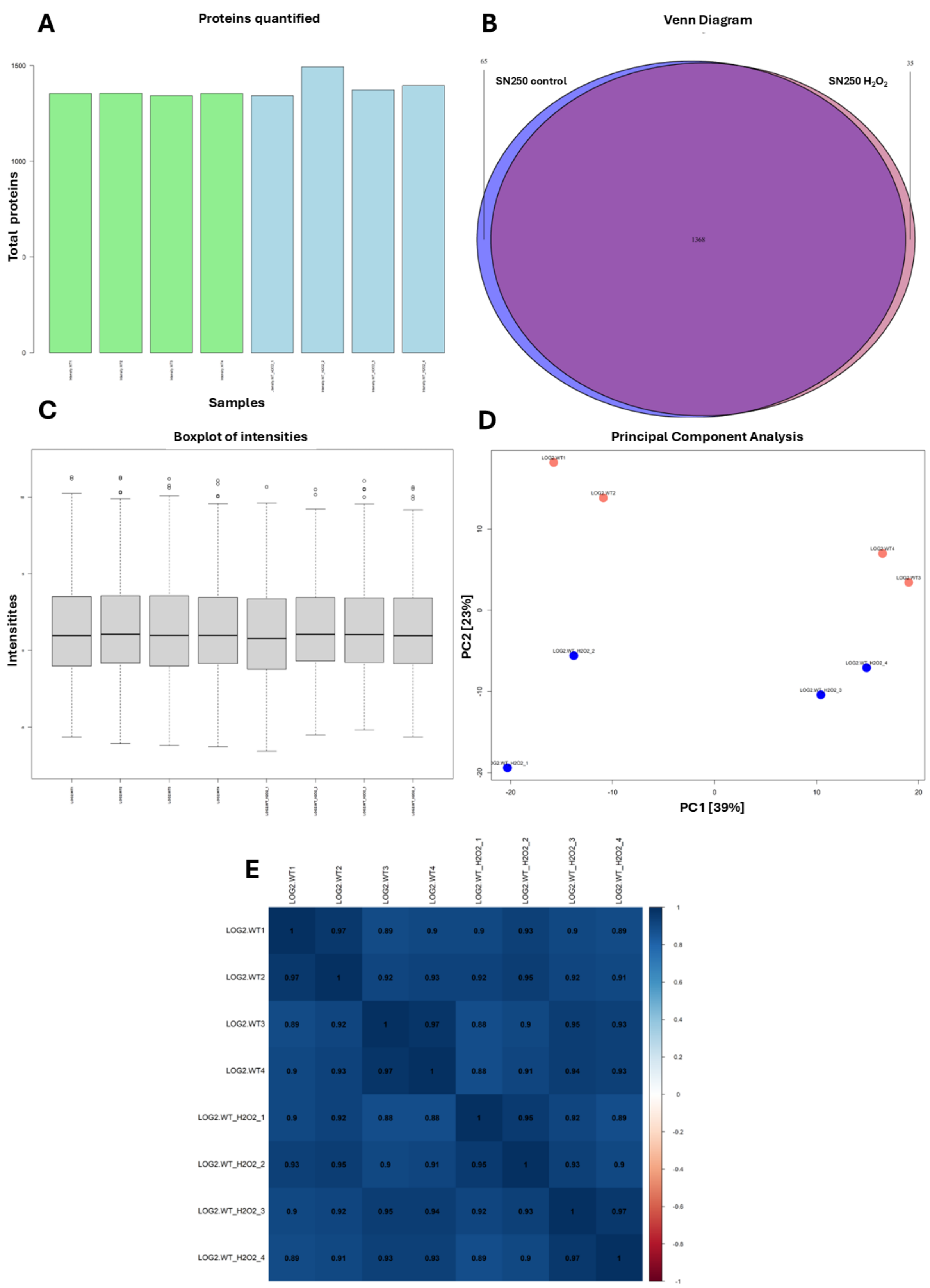

**Figure 2** Illustration of some pre-processing and quality control plots from the TraianProt interface in the SN250 $H_2O_2$ vs SN250 comparison. A) Bar plot of the number of quantified protein groups across biological samples between SN250 $H_2O_2$ and SN250 (control) conditions. B) Venn diagram showing

common and non-common proteins in the two groups compared. C) Boxplot of intensities of the samples belonging to conditions SN250 $H_2O_2$ and SN250 (control). D) Principal Component Analysis (PCA) of the samples belonging to conditions SN250 $H_2O_2$ (dark blue) and SN250 control (salmon). E) Correlation between samples belonging to conditions SN250 $H_2O_2$ and SN250 (control).

Statistical analysis of the relative quantification (two sample T-test *p*-value < 0.01 by limma and PSMs correction applied with DEqMS package) and $log_2FC \geq 0.585$ and $log_2FC \leq -0.585$ revealed that 165 proteins significantly changed their abundance in response to the treatment. It should be noted that the same parameters used in [16] were applied to this analysis in order to enable future comparison. These proteins include Oye22, Prn1, Alt1, Icl1, Orft19.3477 and Cip1, which show a significant increase in relative abundance as stated in [16]. These significant proteins are represented in the volcano plot (Figure 3A) and heatmap (Figure 3B). It should be noted that proteins that are quantified exclusively in one of the studied conditions are considered to be differentially abundant proteins, and the $log_2$ fold-change and p-value are both imputed with the lowest value in the distribution. The entire table can be found in Supplementary file 2 where users can inspect the output format of a differential analysis table from TraianProt.

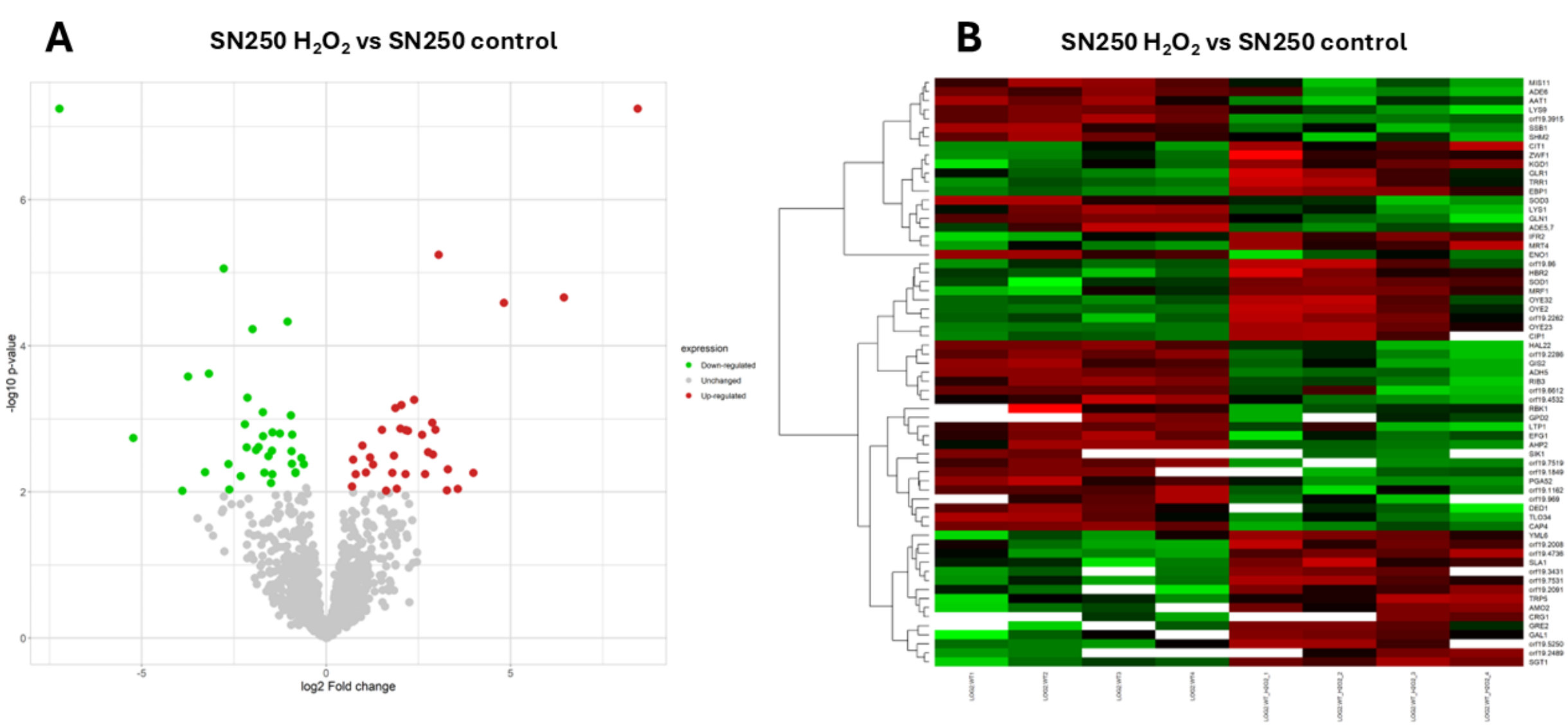

**Figure 3** Statistical analysis plots. A) Volcano plot of significantly abundant proteins in the SN250 $H_2O_2$ vs SN250 comparison B) Heatmap of significantly abundant proteins in the SN250 $H_2O_2$ vs SN250 comparison.

Functional enrichment analysis of the differentially abundant proteins was performed using g:Profiler, by comparing the gene lists of the significantly abundant proteins with a custom background dataset containing all proteins identified during the experiment, specifying the *C. albicans* organism id "calbicans", the Protein ID for the nomenclature conversion as "ENSG" and establishing an enrichment threshold of *p*-value < 0.05, showing a high increase in proteins with the oxidoreductase activity, cell redox homeostasis and NADP(H) oxidoreductase activity GO terms as it is stated in [16]. The most significant GO terms can be visualized in Figure 4A and Figure 4B and the table containing all the GO terms can be found in Supplementary file 3 where users can inspect the output format of a functional analysis table from TraianProt. Moreover, the interaction analysis using the STRINGdb package was performed, specifying *C. albicans* taxon id: “237561” and a score threshold of 0.4. The interaction networks can be visualized in Figure 4C and Figure 4D.

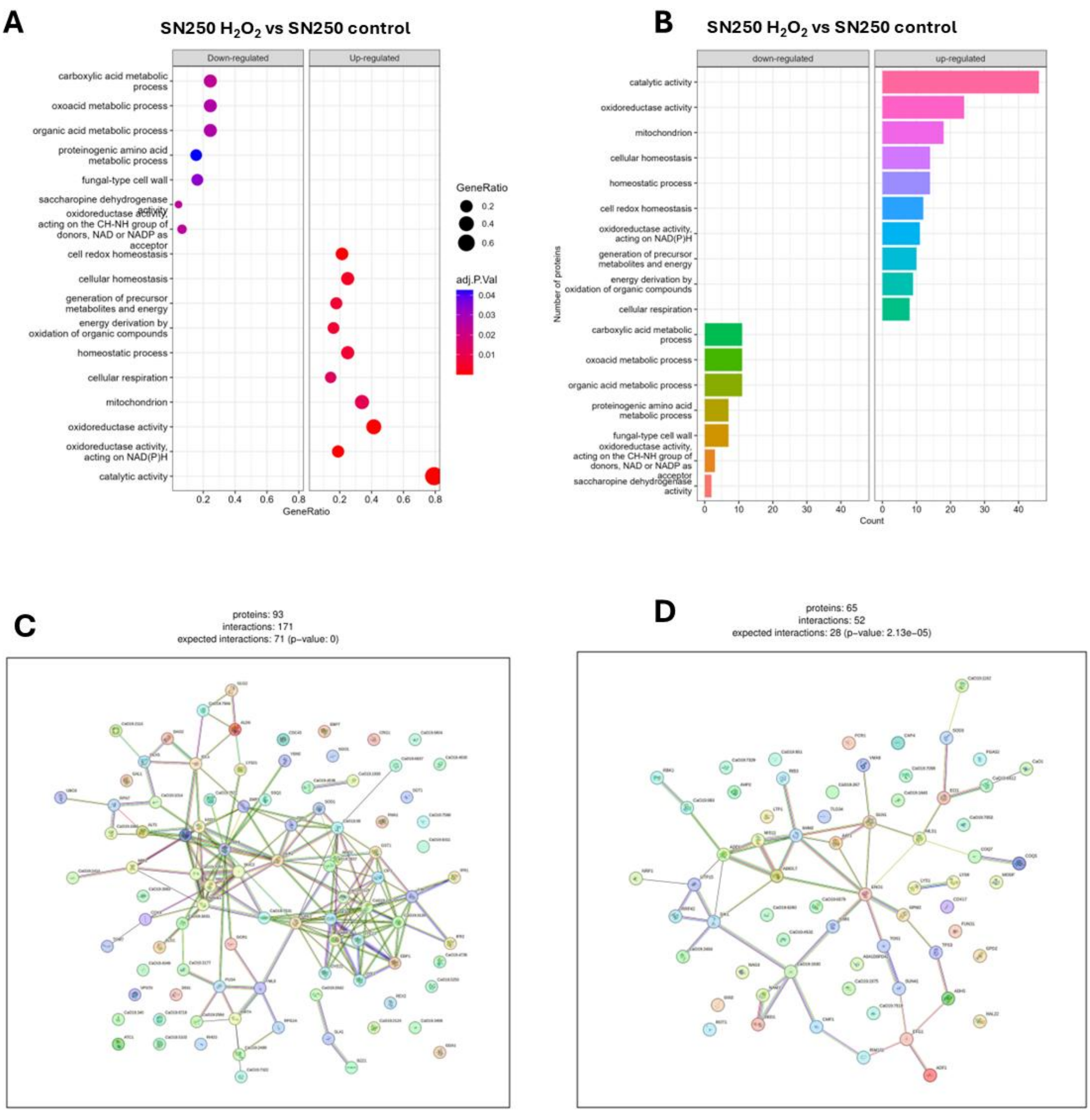

**Figure 4** Functional and interaction analysis plots. A) Dot plot of top significantly Gene ontology enriched terms (Biological Process, Cellular Component and Molecular Function). B) Bar plot of top Gene ontology significantly enriched terms (Biological Process, Cellular Component and Molecular Function). C) Interaction network of up-regulated proteins. D) Interaction network down-regulated proteins.

### 3.3 Use study 2: DIA-NN dataset

The example data was taken from a label-free Data-Independent Acquisition proteomic analysis [18]. In this study the potential for local ribosome maturation in rat neurons is investigated detecting an abundance of ribosome biogenesis factors (RBFs) in distal neuronal compartments, particularly those associated with the late stages of ribosome assembly. These findings challenge conventional models that confine ribosome biogenesis to nuclear and perinuclear regions suggesting that neurons may fine-tune local protein synthesis by regulating ribosome assembly near synaptic sites. The raw files were deposited in the ProteomeXchange Consortium via PRIDE partner repository with the dataset identifier PXD058169. This includes 8 conditions, 4 for cell lysates and 4 for sucrose cushion, 5 replicates in case of untreated conditions and 4 replicates in case of treated conditions.

DIA raw files were processed with the open-source software DIA-NN (version 1.8.2) using a library-free approach. The predicted library was generated using the *in silico* FASTA digest (Trypsin/P) option with the UniProtKB database (Proteome_ID: UP000002494) for *Rattus norvegicus*. The covered peptide length range was set to 7-30 amino acids, missed cleavages to 2 and precursor charge range to 1-5. Methionine oxidation and protein N-terminal acetylation were set as variable modifications. Cysteine carbamidomethylation was selected as a fixed modification. The maximum number of variable modifications per peptide was limited to 1. MS1 and MS2 accuracies as well as scan-windows were set to 0, isotopologues were enabled, while match-between-runs and shared spectra were disabled [18].

The proteomic analysis enabled the quantification of 8872 *R. norvegicus* proteins in the "lysate somata + neurites vs neurites only" comparison. Protein groups were filtered for 1 % False Discovery Rate (FDR) at protein and precursor level. The intensity values of the quantified proteins were transformed into a logarithmic scale. Only common protein groups with intensity values in at least 50% of the samples are included in the analysis. Furthermore, a protein group is retained when at least one unique peptide has been identified in a proportion of 50 % of the samples in at least one of the two groups being compared. This filter was applied using the local TraianProt version, from which the unique peptide information was subtracted from the 'report.tsv' file. The normalization method applied was median normalization; protein intensities were normalized by subtracting the median for each biological replicate. No imputation method was selected; missing values will be handled in steps taken downstream (Figure 5).

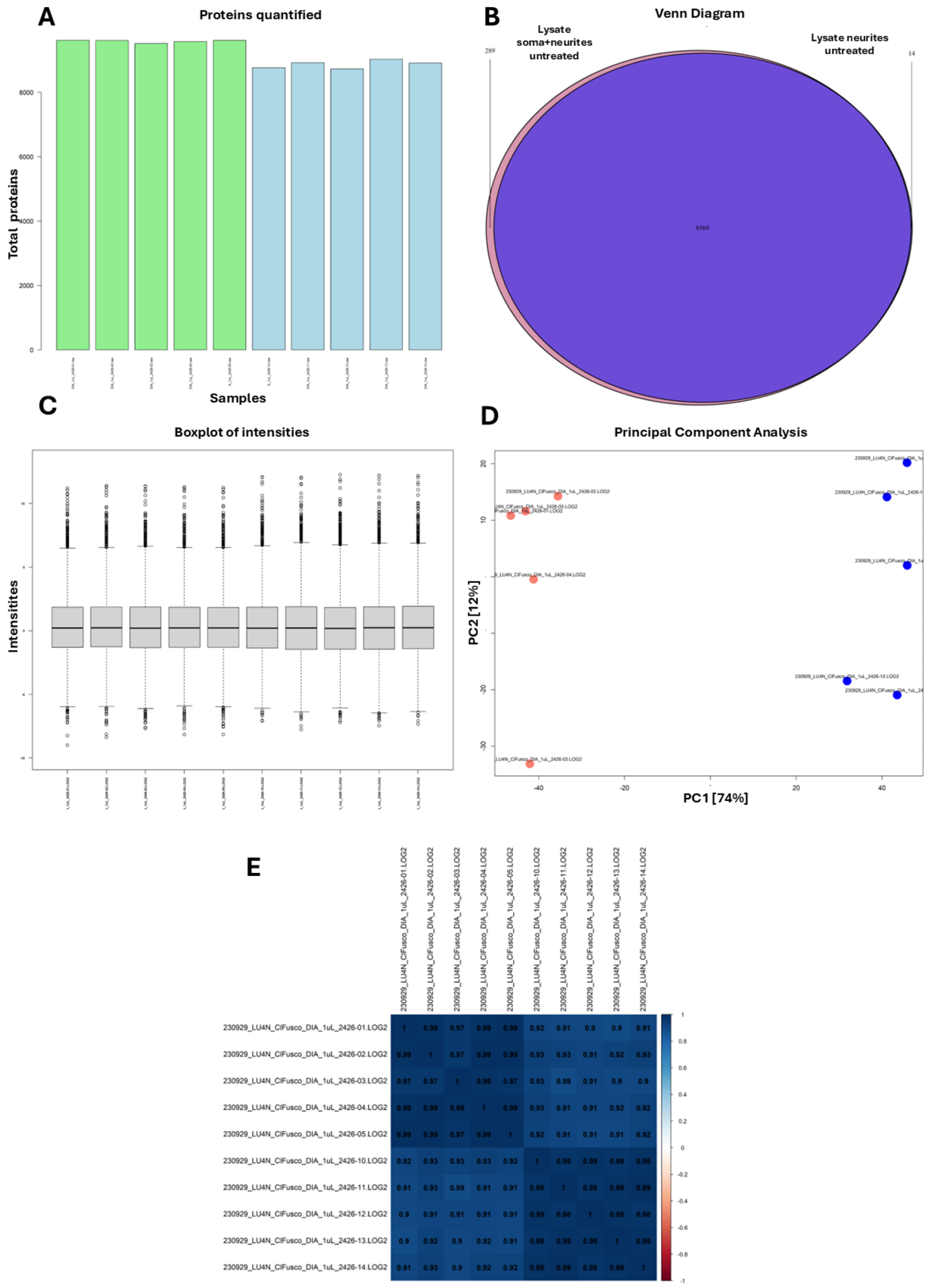

**Figure 5** Illustration of some pre-processing and quality control plots from the TraianProt interface in the lysate somata + neurites vs neurites only comparison. A) Bar plot of the number of quantified protein groups across biological samples between lysate somata + neurites and neurites only conditions. B) Venn diagram showing common and non-common proteins in the two groups compared. C) Boxplot of intensities of the samples belonging to lysate somata + neurites and neurites only conditions D) Principal Component Analysis (PCA) of the samples belonging to lysate somata + neurites (dark blue) and neurites only conditions (salmon). E) Correlation between samples belonging to lysate somata + neurites and neurites only conditions.

Statistical analysis of the relative quantification (two sample T-test *p*-value < 0.01 by limma and PSMs correction applied with DEqMS package) and $log_2FC \geq 1$ and $log_2FC \leq -1$ revealed that 2030 proteins significantly changed their abundance in response to the treatment. It should be noted that the same parameters used in [18] were applied to this analysis in order to enable future comparison. These significant proteins are represented in the volcano plot (Figure 6A) and heatmap (Figure 6B). It should be noted that proteins that are quantified exclusively in one of the studied conditions are considered to be differentially abundant proteins, and the $log_2$ fold-change and p-value are both imputed with the lowest value in the distribution. The complete list can be found in Supplementary file 4.

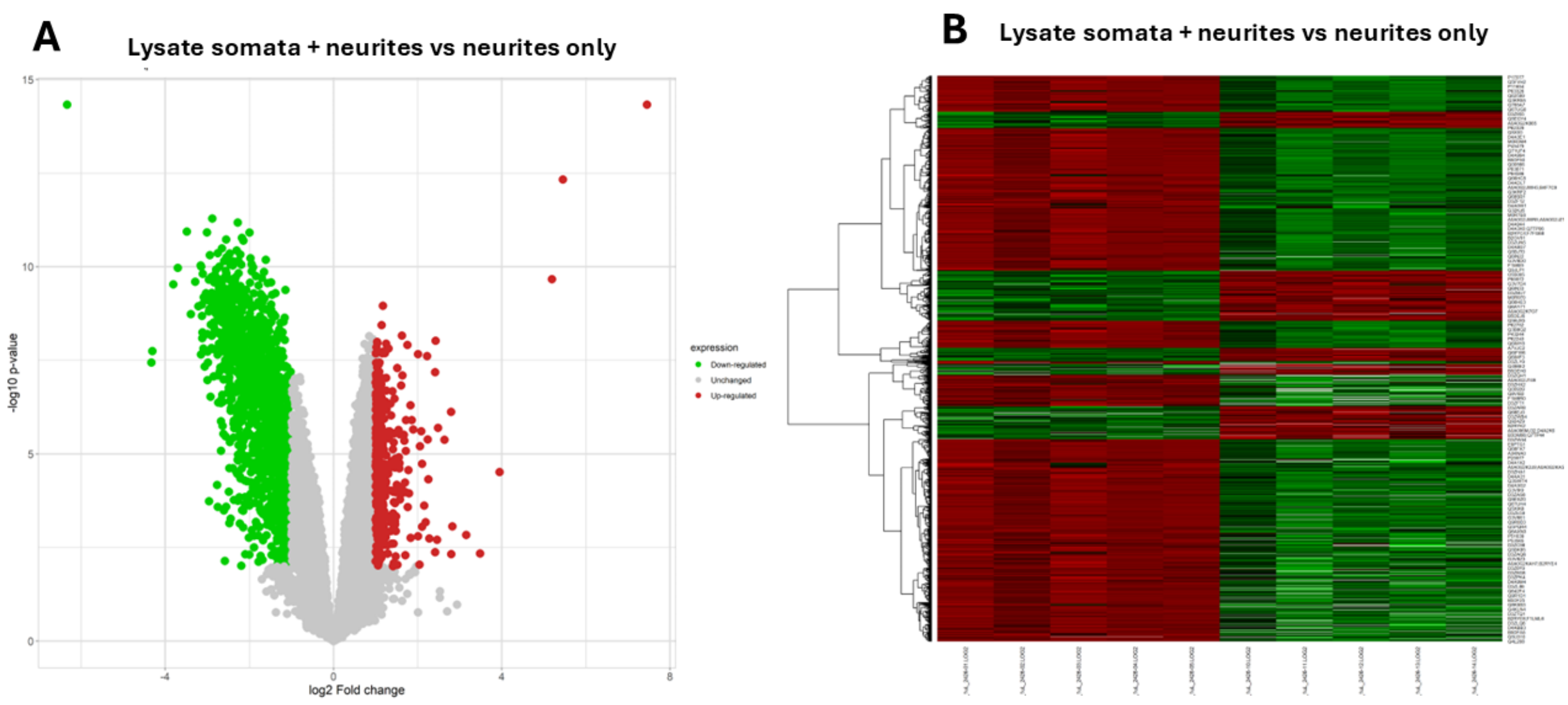

**Figure 6** Statistical analysis plots. A) Volcano plot of significantly abundant proteins in the lysate somata + neurites vs neurites only comparison. B) Heatmap of significantly abundant proteins in the lysate somata + neurites vs neurites only comparison.

Functional enrichment analysis of the differentially abundant proteins was performed using g:Profiler, by comparing the gene lists of the significantly abundant proteins with a custom background dataset containing all proteins identified during the experiment, specifying the *R. norvegicus* organism id "rnorvegicus", the Protein ID for the nomenclature conversion as "ENSG" and establishing an enrichment threshold of *p*-value < 0.05. The most significant GO terms can be visualized in Figure 7. A decrease in the abundance of proteins related with the spliceosomal complex, the regulation of mRNA or the Endoplasmic reticulum lumen is shown as it is stated in [18]. The complete table with the GO terms can be found in Supplementary file 5. Moreover, the interaction analysis using the STRINGdb package was performed, specifying *R. norvegicus* taxon id: "10116" and a score threshold of 0.4. The interaction networks can be visualized in Figure 7C and Figure 7D.

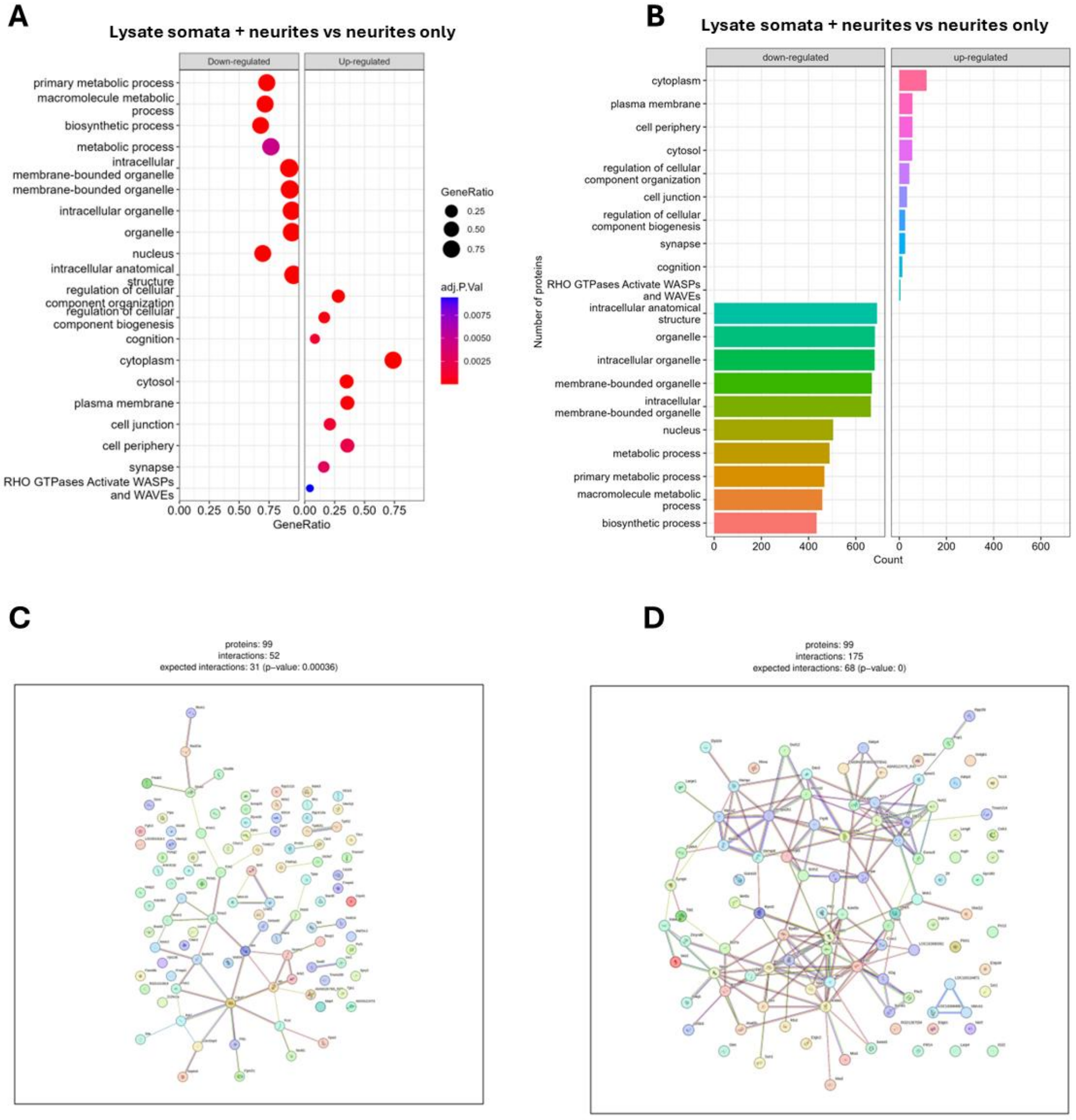

**Figure 7** Functional analysis plots. A) Dot plot of top significantly Gene ontology enriched terms (Biological Process, Cellular Component and Molecular Function). B) Bar plot of top Gene ontology significantly enriched terms (Biological Process, Cellular Component and Molecular Function). C) Interaction network of up-regulated proteins. D) Interaction network down-regulated proteins.

## Comparison with other tools

The analysis of large-scale proteomic data typically requires the use of multiple independent tools and software packages to address different downstream tasks. Nonetheless, there is an increasing demand for integrative, web-based platforms that consolidate these functionalities into a single interface, thereby facilitating streamlined and sequential data analysis, particularly for researchers without programming expertise. TraianProt has been specifically developed to address this limitation by providing an intuitive platform for the analysis of complete proteome datasets derived from both labeled and label-free experiments, involving rigorous data filtering by both intensity values and number of unique peptides, normalization, and imputation, differential expression testing with PSM correction, as well as enrichment and interaction analyses across a wide range of organisms. Importantly, the platform integrates all these functions for both DDA and DIA-based datasets and can be accessed through the web server version, local version or the R package version. Table 1 provides a comparative benchmark of TraianProt against other leading, state-of-the-art platforms for downstream proteomics analysis. Some of the most robust tools offer access to their software in various ways, such as via an online web server, an R package or locally, such as Phosmap [10], Tidyproteomics [19] and ProStaR [7]. However, this benchmark reveals critical gaps in the analytical capabilities of the current software landscape. While most platforms cover standard workflows like normalization and differential expression, they often lack comprehensive, integrated modules for deeper biological interpretation. For instance, prominent tools like Perseus [1], Provision [3], and Fragpipe analyst [4] lack functional analysis modules that cover a wide range of organisms, while others, including ProStaR [7]and Tidyproteomics [19], omit protein-protein interaction network analysis. More critically, our comparison highlights a general absence of the unique peptide filtering methods required for high-confidence downstream analysis. TraianProt is the only platform among those benchmarked that provides a method for PSM-level correction, a crucial step for accurate variance estimation. Furthermore, it is one of only two tools (along with Provision) to implement a stringent unique peptide filter which assesses the biomarker nomination, the difference is that Provision applies this filter based on the total number of unique peptides identified per protein group, whereas TraianProt considers the number of unique peptides identified in a user defined minimum proportion of samples in at least one experimental condition.

**Table 1** Comparison of TraianProt with several state-of-art proteomics downstream analysis tools.

| Characteristics | | Tools | | | | | | | | |
|---|---|---|---|---|---|---|---|---|---|---|
| | | Perseus | Provision | Fragpipe analyst | MCQR | Amica | Phosmap | Tidyproteomics | ProStaR | TraianProt |
| Acquisition method | DDA | ✓ | ✓ | ✓ | ✓ | ✓ | ✓ | ✓ | ✓ | ✓ |
| | DIA | ✓ | ✗ | ✓ | ✓ | ✓ | ✓ | ✓ | ✓ | ✓ |
| Data pre-processing | Support various search engines | ✗ | ✗ | ✓ | ✓ | ✓ | ✓ | ✓ | ✓ | ✓ |
| | Unique peptides filter | ✗ | ✓ | ✗ | ✗ | ✗ | ✗ | ✗ | ✗ | ✓ |
| | Normalization | ✓ | ✓ | ✓ | ✓ | ✓ | ✓ | ✓ | ✓ | ✓ |
| | Filtering | ✓ | ✓ | ✓ | ✓ | ✓ | ✓ | ✓ | ✓ | ✓ |
| Downstream analysis | Dimension reduction analysis | ✓ | ✓ | ✓ | ✓ | ✓ | ✓ | ✓ | ✓ | ✓ |
| | Differential expresión analysis | ✓ | ✓ | ✓ | ✓ | ✓ | ✓ | ✓ | ✓ | ✓ |
| | PSM correction | ✗ | ✗ | ✗ | ✗ | ✗ | ✗ | ✗ | ✗ | ✓ |
| | Functional analysis covering a wide range of organisms | ✗ | ✗ | ✗ | ✗ | ✓ | ✓ | ✓ | ✓ | ✓ |
| | STRING | ✓ | ✓ | ✗ | ✗ | ✓ | ✓ | ✗ | ✗ | ✓ |
| Platform compatibility | Windows | ✓ | ✓ | ✓ | ✓ | ✓ | ✓ | ✓ | ✓ | ✓ |
| | Mac OS X | ✗ | ✓ | ✓ | ✓ | ✓ | ✓ | ✓ | ✓ | ✓ |
| | Linux | ✗ | ✓ | ✓ | ✓ | ✓ | ✓ | ✓ | ✓ | ✓ |
| Web server | On-line interactive analysis | ✗ | ✓ | ✓ | ✓ | ✓ | ✓ | ✓ | ✓ | ✓ |
| R package | Bioconductor R package | ✗ | ✗ | ✗ | ✗ | ✗ | ✓ | ✓ | ✓ | ✓ |

To evaluate the performance of TraianProt, we compared it with established proteomics analysis pipelines. Firstly, we reprocessed the .xlsx file containing quantified protein groups from [16] using TraianProt selecting the same parameters and contrasted the resulting fold-change estimates and associated p-values with those obtained through differential expression analysis in Proteome Discoverer (v2.4). This direct comparison allowed us to assess the ability of TraianProt to reproduce results generated in this case by widely adopted commercial software. To visualize the degree of concordance, we examined the overlap of significantly regulated proteins Figure 8A between the two approaches and generated scatterplots comparing fold-change Figure 8B and p-value Figure 8C estimates. An overlap was observed between the two downstream analysis approaches, with 135 out of 353 proteins (38%) identified as differentially expressed by both. In addition to this overlap, Proteome Discoverer reported 102 proteins as significant that did not pass the p-value threshold when analyzed with TraianProt. This discrepancy can be attributed to several factors. First, the raw data were processed with different search engines: in [16], Sequest HT (Thermo Scientific, Waltham, MA, USA) was employed, whereas in our analysis MaxQuant was used. Such differences in search algorithms inherently affect peptide identification rates, which in turn influence the number of quantified proteins that reach statistical significance. Furthermore, the statistical framework implemented in Proteome Discoverer differs from that of limma, which employs an empirical Bayes approach to moderate variance estimates across protein. Conversely, TraianProt identified 116 proteins as significant that were not detected by Proteome Discoverer. This may reflect the variance shrinkage of limma in detecting subtle yet consistent expression changes across replicates, which can enhance power in datasets with limited sample size or heterogeneous variance structures. Importantly, for the subset of proteins consistently identified as significant by both approaches, fold changes and associated p-values were highly concordant, underscoring the robustness of the shared findings.

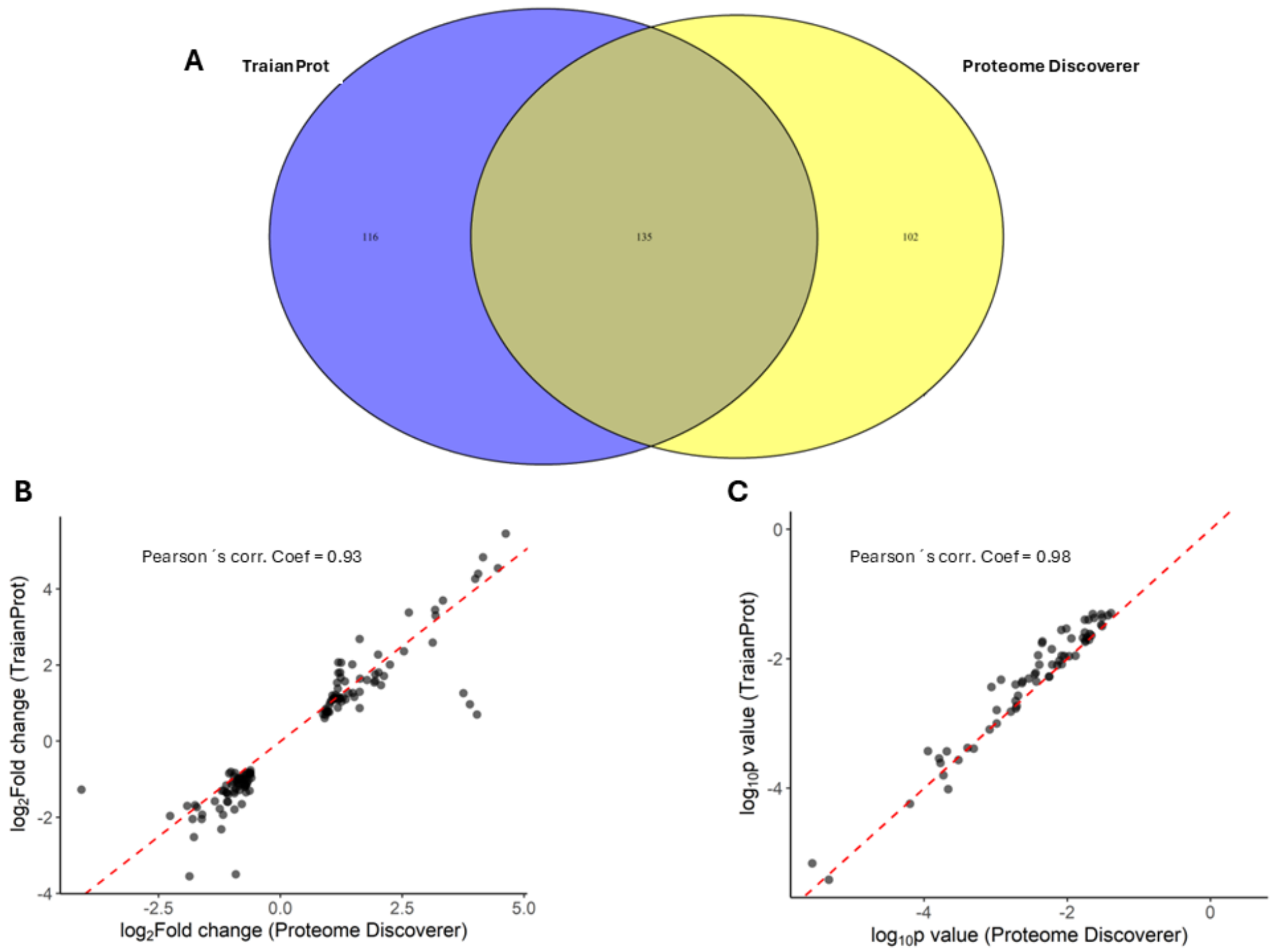

**Figure 8** Benchmarking of TraianProt against Proteome Discoverer. A) Venn diagram showing the overlap of the differentially expressed proteins reported by TraianProt and Proteome Discoverer v2.4.

B) Scatterplot of the resulting protein $\log_2$ fold changes from the differentially expressed proteins as reported by TraianProt and Proteome Discoverer v2.4. C) Scatterplot of the resulting $\log_{10}$ p values from the differentially expressed proteins as reported by TraianProt and Proteome Discoverer v2.4.

Secondly, we reprocessed the report.pg_matrix.tsv file containing quantified protein groups from [18] using TraianProt selecting the same parameters and contrasted the resulting fold-change estimates and associated p-values with those obtained through DIA-NN R package and the in-house script developed by the authors. To visualize the degree of concordance, we again compared the results obtained by TraianProt to those derived from the DIA-NN R package and the in-house script developed by the authors. The overlap of significantly regulated proteins between the two approaches Figure 9A and generated scatterplots comparing fold-change Figure 9B and p-value Figure 9C estimates. A substantial overlap was observed with 1349 (out of 2196) proteins (61%) being identified as differentially expressed. Beyond this common set, the in-house script reported 440 additional significant proteins that were not retained by TraianProt, most likely due to differences in the statistical treatment of variance and significance thresholds. Conversely, TraianProt detected 408 proteins as significant that were not identified by the in-house script, reflecting the empirical Bayes framework implemented in limma. Despite these differences, the proteins consistently identified as differentially expressed by both approaches displayed highly similar fold changes and associated significance values, reinforcing the robustness of the shared findings.

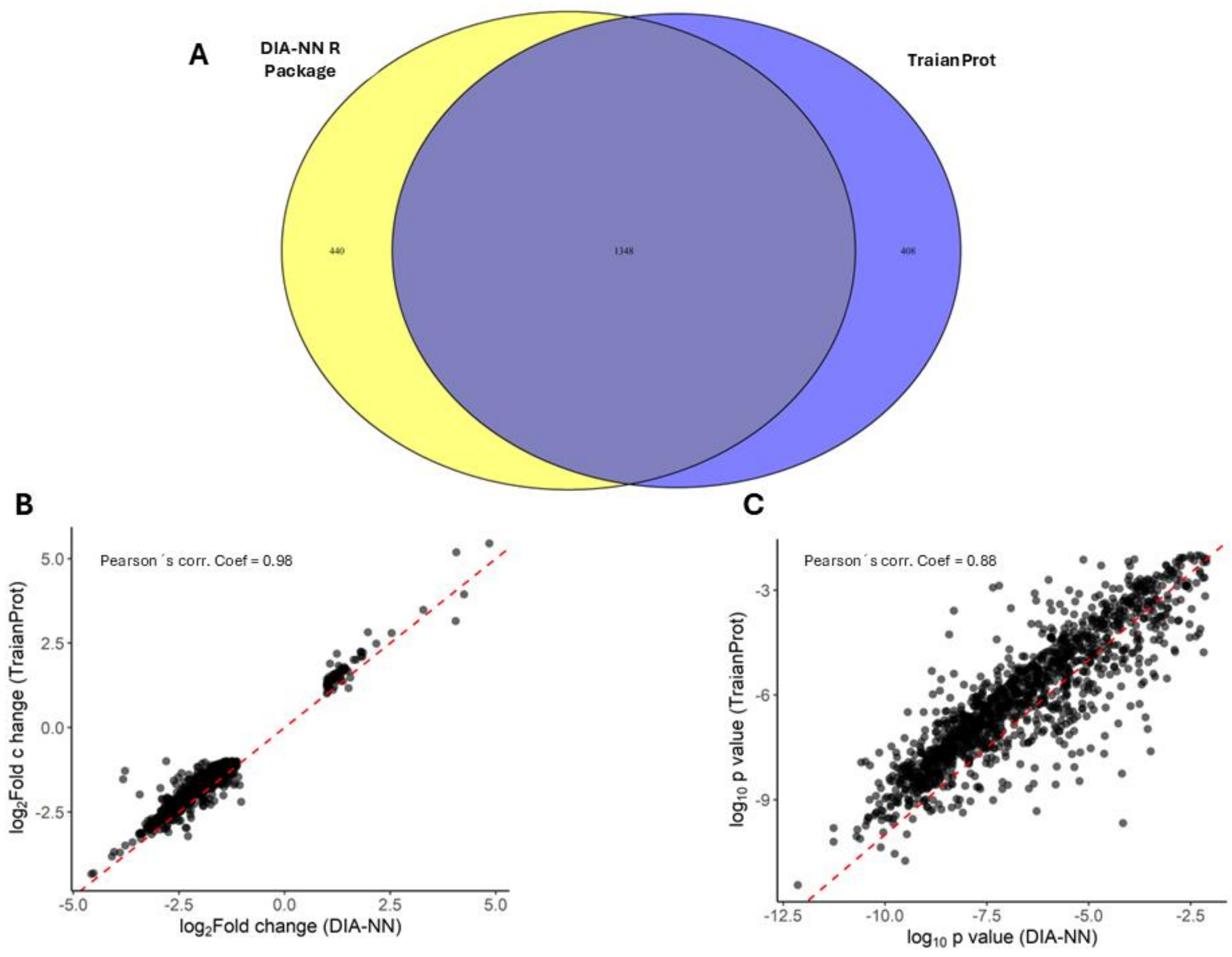

**Figure 9** Benchmarking of TraianProt against DIANN R package. A ) Venn diagram showing the overlap of the differentially expressed proteins reported by TraianProt the DIANN R package. B) Scatterplot of the resulting protein $\log_2$ fold changes from the differentially expressed proteins as reported by TraianProt and the DIANN R package. C) Scatterplot of the resulting $\log_{10}$ p values from the differentially expressed proteins as reported by TraianProt and Proteome Discoverer the DIANN R package.

Taken together, the differential expression analyses conducted using TraianProt, the base R DIA-NN package, and Proteome Discoverer v2.4 produced very similar results. In particular, the calculated fold changes and p-values were highly similar, which clearly validates the analysis pipeline implemented in TraianProt.

## Conclusions and future perspectives

TraianProt is an open-source web-based, user-friendly platform designed for proteomics data analysis supporting output formats from leading computational platforms. Its primary aim is to facilitate the extraction of interpretable results from proteomic data by generating high-quality, customizable graphs and statistics in a dynamic environment. The platform allows users to revisit the analysis and adjust parameters to achieve the most optimal outcomes and makes it accessible for new users by not requiring any programming knowledge. Furthermore, TraianProt incorporates novel filtering strategies, such as the unique peptide filter, which enhances the reliability and robustness of the results. This platform is freely available in the following Github repository: https://github.com/SamueldelaCamaraFuentes/TraianProt

Nevertheless, some considerations warrant further discussion. As with any integrative platform, continuous updates will be necessary to ensure compatibility with newly emerging search engines. For example, while TraianProt already supports a range of upstream tools including MaxQuant, MSFragger, DIA-NN, Proteome Discoverer and ProteoScape, future work must ensure that it remains compatible with new or updated versions of these tools, as well as with novel search engines or file output formats that may adopt different data schemas. Also, because TraianProt is web-based, issues of, user upload size limits, server uptime, and scalability for multi-user environments or with very large datasets should be addressed explicitly in future versions.

## Funding

This work was granted by PID2021-124062NB-I00 and PID2024-156142NB-I00 funded by Spanish Ministry of Science and Innovation/ State Research Agency 10.13039/501100011033. S.C.F. was supported by PTA2022-022414-I from the Spanish Ministry of Science and Innovation and the State Research Agency, along with REACT-UE ANTICIPA-CM from the Community of Madrid.

## Acknowledgements

Not applicable

## Author contributions

S.C.F wrote the main manuscript text, prepared figures, designed the work, created the software and interpreted data presented in this work, D.G.B, designed the work, interpreted data, prepared figures and substantively revised the work , M.L.H designed the work, interpreted data and substantively revised the work C.G designed the work, interpreted data and substantively revised the work.All authors reviewed the manuscript.

## Availability and requirements

**Project name:** TraianProt
**Project home page:** https://github.com/SamueldelaCamaraFuentes/TraianProt
**Operating system(s):** e.g. Platform independent
**Programming language:** RStudio
**Other requirements:** In case it is used through the web server: nothing else

**License:** GNU General Public License.
**Any restrictions to use by non-academics:** None

**Availability of data and materials**

Source code is available from https://github.com/SamueldelaCamaraFuentes/TraianProt and the software can be accessed through https://github.com/SamueldelaCamaraFuentes/TraianProt. The unique peptides extractor tool is available at https://samueldelacamara.shinyapps.io/Unique_peptides_extractor/ which can also be accessed through TraianProt. A PDF version of the tutorial is provided in the supplementary data. Data used in the use study example analysis is available in https://github.com/SamueldelaCamaraFuentes/TraianProt/test which were obtained from PRIDE whose dataset identifier is PXD040804 for use study 1 and PXD058169 for use study 2. The database used for the use study 1 Is publicly available in the Candida Genome Database: http://www.candidagenome.org/download/sequence/C_albicans_SC5314/Assembly22/current/ .

**Declarations**

**Ethics approval and consent to participate**

Not applicable

**Ethics declaration**

Not applicable

**Consent for publication**

Not applicable

**Competing interests**

The authors declare that they have no competing interests

**List of abbreviations**

**TMT:** Tandem Mass Tag**.**

**SILAC:** Stable Isotope Labeling with Amino Acid in Cell Culture.

**DDA:** Data Dependent Acquisition.

**DIA:** Data Independent Acquisition.

**PSM:** Peptide-Spectrum Match**.**

**FDR:** False Discovery Rate.

**VSN:** Variance Stabilization Normalization.

**CGD:** Candida Genome Database**.**

**MBR**: Match Between Runs**.**

**PCA:** Principal Component Analysis**.**

**PRIDE:** Proteomics Identifications Database.